# How Lattice and Charge Fluctuations Control Carrier Dynamics in Halide Perovskites

Matthew Z. Mayers,[1] Liang Z. Tan,[2] David A. Egger,[3] Andrew M. Rappe,[2] and David R. Reichman[1]

[1]Department of Chemistry, Columbia University, New York, New York, 10027, USA

[2]Department of Chemistry, University of Pennsylvania, Philadelphia, Pennsylvania, 19104-6323, USA

[3]Institute of Theoretical Physics, University of Regensburg, 93040 Regensburg, Germany

## Abstract

Here we develop a microscopic approach aimed at the description of a suite of physical effects related to carrier transport in, and the optical properties of, halide perovskites. Our theory is based on the description of the nuclear dynamics to all orders and goes beyond the common assumption of linear electron-phonon coupling in describing the carrier dynamics and band gap characteristics. When combined with first-principles calculations and applied to the prototypical MAPbI$_3$ system, our theory explains seemingly disparate experimental findings associated with both the charge-carrier mobility and optical absorption properties, including their temperature dependencies. Our findings demonstrate that orbital overlap fluctuations in the lead-halide structure plays a significant role in determining the optoelectronic features of halide perovskites.

**lattice fluctuations & charge fluctuations**

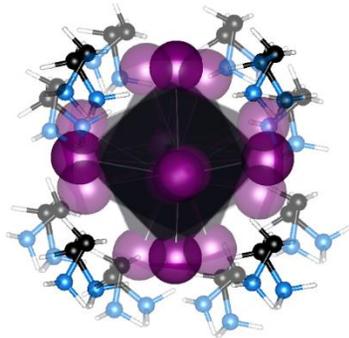
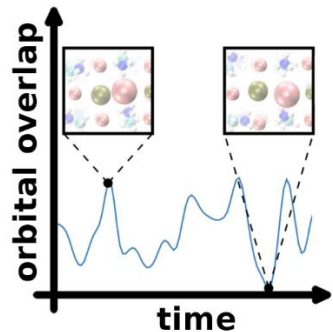

Halide perovskites (HaPs) such as MAPbI$_3$ have found applications in several optoelectronic devices[1–5] including photovoltaic cells, where power conversion efficiencies rival those of highly-processed crystalline silicon despite a solution-based route to their synthesis.[6–9] High efficiencies are related to long carrier diffusion lengths, as dictated by the carrier mobility-lifetime product, and near-optimal absorption features, including a sharp absorption edge.[10] These advantageous properties are likely linked to the unusual physical characteristics of the lead-halide bonds, notably their mixed ionic-covalent character and their low-frequency, large-amplitude anharmonic lattice displacements, which generate significant polar fluctuations.[11,12] Concomitantly, the HaP electronic band-structure is reflective of typical inorganic semiconductors, with low effective masses of holes and electrons, suggesting large mobilities and efficient carrier transport. This combination of nuclear and electronic properties appears to be unique in the context of high-efficiency semiconductors.

The unusual nature of lattice fluctuations in HaPs renders traditional solid-state approaches, which generally rely on small (harmonic) displacement models with linear electron-phonon coupling,[13,14] suspect for the description of the carrier and optical absorption properties of these compounds. It has been shown that conventional theories either generally strongly underestimate the temperature ($T$) dependence of HaP mobilities or misrepresent their room-$T$ magnitudes.[15–17] For example, acoustic-phonon scattering models yield a room-$T$ value that is several orders of magnitude above experimental data,[15] while large polaron theories based on optical-mode scattering strongly underestimate its $T$-dependence.[16–19] A very recent *ab initio* study employing linear coupling to optical modes was successful in describing the $T$-dependence of the carrier lifetimes in MAPbI$_3$.[20] With respect to this latter study, it should be noted that while models based on the assumption of low-order electron-phonon coupling form the standard basis for the description of the $T$-dependence of semiconductor band gaps,[21,22] such models appear not to capture the $T$-dependence of the MAPbI$_3$ band gap.[23,24] In particular, a proper description of this effect was found to require strongly non-linear electron-phonon coupling.[25] Lastly, there is currently no microscopic description of how the large-amplitude anharmonic lattice displacements in HaPs, which seemingly should yield large fluctuations in the band gap, relate to the sharp optical absorption edge of MAPbI$_3$ at room-$T$.[10] To the best of our knowledge, a unified microscopic theory that would rationalize all of these important physical characteristics of HaPs does not currently exist.

Surmounting the challenges presented by these puzzles will thus greatly enable the design of new materials that build upon the attractive features of HaPs.

Here, we take a major step in this direction via the formulation and theoretical investigation of a fully microscopic model describing the interplay between electronic structure and nuclear motion in MAPbI$_3$ which makes no limiting assumptions about the harmonic nature of the lattice or the specifics of the electron- (hole-) phonon coupling. When our model is combined with first-principles calculations for the prototypical HaP variant MAPbI$_3$, it provides a reasonably accurate description of the charge-carrier mobility and optical absorption properties, including their *T*-dependencies, as validated by comparison to pertinent experimental data. The theory highlights the importance of anharmonic fluctuations of the Pb-I bonds that evolve in time and space and produce large temporal and spatial modulations of orbital overlaps. In contrast to canonical inorganic semiconductors, the charge carrier dynamics in MAPbI$_3$ are thus characterized by sizable off-diagonal electron-phonon coupling effects, which limit carrier mobilities without resulting in substantial carrier trapping or impacting the sharpness of the absorption edge in MAPbI$_3$.

Our model aims to simulate very large system sizes and thus begins with a computationally efficient tight-binding (TB) parameterization of the band structure, as previously done, see, e.g., Refs. [26,27]. It includes real-space descriptions of the on-site energies associated with occupancy of the relevant s-and p-orbitals of Pb and I, as well as the kinetic energy arising from π and σ overlaps of these orbitals, which determine the hopping parameters in the TB scheme. We note that our TB scheme proceeds via a projection of the Bloch wavefunctions, calculated using density functional theory (DFT), onto atomic wavefunctions as described in the Supporting Information (SI). The TB approach thus naturally includes the important effect of spin-orbit coupling taken into account at the DFT level. Furthermore, the underlying PBE exchange-correlation functional was found to provide a fairly accurate description of the effective masses of MAPbI$_3$, which are important for the theoretical evaluation of the carrier dynamics. Orbital overlaps and on-site energies are functions of the instantaneous nuclear configuration and may be updated as the nuclei move at each new set of lattice locations. As the lattice motion in HaPs exhibits anharmonic contributions around room-*T*, in our microscopic model we stipulate to go *beyond the harmonic approximation and describe the nuclear displacements to all orders* by virtue of fully unconstrained molecular dynamics (MD). These may be straightforwardly generated from a given force field[28] or via *ab-* *initio* molecular dynamics (AIMD) whereby the force on each atom is generated on-the-fly from the instantaneous electronic density calculated self-consistently in DFT. Here we select the former approach, as the latter is currently computationally infeasible for the subsequent quantum dynamics simulations. The force field used here is known to reproduce *T*-dependent dynamical changes of the MAPbI$_3$ lattice and stands in reasonable agreement with AIMD calculations.[29] Due to the fact that it produces an elastic modulus that is approximately 20% too *large*,[28] it perhaps slightly *underestimates* the size of the fluctuations highlighted below. Further details can be found in the SI.

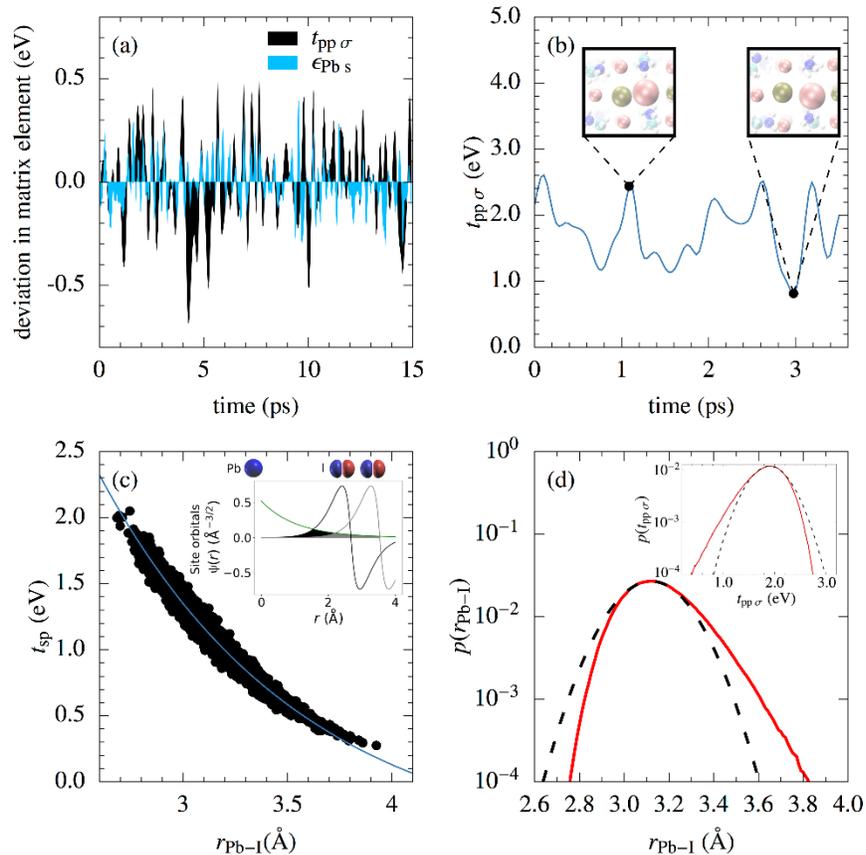

**Figure 1** (a) Typical dynamical fluctuations in MAPbI$_3$ at 300 K of orbital overlap and on-site energies as calculated via the DFT-tight-binding approach. The dynamical fluctuations in the orbital overlap generally exceed those in the on-site energies, which is unique for polar semiconductors. (b) Fluctuations of the orbital overlap, illustrating how large changes in the lead (gold)-iodine (pink) distance induce large fluctuations. (c) An exponential fit of the distance dependence of one of the hopping terms; the inset depicts a sketch of the change in Pb-I orbital overlap corresponding to the two highlighted points in panel (b). (d) Histogram (red curves) of the dynamically changing Pb-I distances and of the orbital overlap (shown in the inset), both recorded at 300 K. The distributions of both observables show strong deviations from a Gaussian behavior (dashed curves).

Fig.1a shows the dynamical fluctuations of both the orbital overlap and on-site energy for a particular bond and atom type when the DFT-based tight binding parameters are recomputed every time step along a representative MD trajectory of nuclear motion around room-*T*. We note that the fluctuations in the on-site energies of MAPbI$_3$ on the basis of a TB model were highlighted before.[30] Strikingly, here we find that the scale of the fluctuations in the orbital overlap is sizable (on the order of 15-20% of the mean value of the hopping parameter) and generically *larger* than that of the on-site energy fluctuations. Fig. 1b shows that the microscopic origin of the fluctuations is rooted in changes of Pb-I covalency due to anomalously large displacements within the mixed ionic-covalent inorganic framework, reminiscent of off-diagonal *intermolecular* fluctuations in organic semiconductors.[31] Unsurprisingly, the orbital overlaps may be well-fit by a simple exponential function of the distance between the Pb and I atoms for all three relevant bond types. Fig.1c shows an exponential fit of the orbital overlap values generated over a series of ~15 ps MD trajectories with the DFT-TB approach, while the inset sketches the Pb-I orbital overlap for short and long Pb-I bonds. The correspondingly large non-linear off-diagonal fluctuations induce scattering mechanisms that, within the family of polar semiconductors, appears unique to HaPs.

When the fluctuations in the orbital overlap and on-site energy TB parameters are expressed in terms of a small-amplitude, linear-order Taylor expansion of the nuclear coordinates, they generate Peierls-type (off-diagonal) and Fröhlich-type (polaronic or diagonal) electron-phonon couplings, respectively.[13,32] Importantly, state-of-the-art first-principles descriptions of electron-phonon interactions in semiconductors usually do not account for Peierls coupling (which imparts a momentum-dependence to the electron-phonon vertex), and secondly typically only consider the linear term in the Taylor expansion of the electron-lattice potential.[13,14] Here, we find that in MAPbI$_3$ the fluctuations in the off-diagonal coupling are large and cannot be ignored. Furthermore, in Fig. 1d we show histograms of the statistics of the fluctuations of the Pb-I bond length and orbital overlap. Both distributions are markedly non-Gaussian, which demonstrates *that Pb-I displacements are anharmonic and that the electron-phonon couplings are effectively non-linear*. A spectral analysis (see SI) shows that the phonon modes which dominate the hopping fluctuations comprise a low frequency band that includes the dominant TO mode as well as motion with characteristic frequencies commensurate with LO modes associated with Pb-I-Pb bending and Pb-I stretching,[20] while the on-site energy fluctuations are controlled by the dominant LO mode.

Having established that anharmonic nuclear motions and off-diagonal, non-linear electron-phonon coupling need to be taken into account when describing carrier dynamics in MAPbI$_3$, our next step concerns examining their effect on quantum transport of electrons and holes. An explicit nuclear-coordinate dependence of the TB parameters, as described in the SI, enables an approximate description of the quantum dynamics in MAPbI$_3$. In particular, a semiclassical (Ehrenfest) treatment,[32] where the electronic degrees of freedom are computed exactly while the nuclear degrees of freedom are treated classically (albeit with no assumptions of small amplitude or harmonic displacement), allows for a facile extraction of the charge mobility,[31] as described in the SI. We have carried out full Ehrenfest calculations including the mean-field back-reaction force of the charge on the nuclei, but we find the effect of this force to be very small (~1% changes to the mobility) similar to results of previous models of charge transport in organic crystals.[33] We thus neglect the back-reaction force for data production runs. We also note that while this approach has known deficiencies,[32] it should be accurate when the frequency of nuclear motion is very small compared to that associated with the magnitude of the mean orbital overlap terms, as is the case in MAPbI$_3$. Treatments which quantize the nuclear motion and go beyond a semiclassical approach as well as numerical details (e.g., initial conditions and trajectory lengths), are reported in the SI.

As shown in Fig. 2a and b, the time evolution of the carrier is affected strongly by the temporal and spatial fluctuations of the orbital overlaps: *charge carriers in MAPbI$_3$ diffuse through a potential energy landscape* (see Fig. 2c) that involves disordered couplings within the lead-halide lattice, which retards a coherent spread of the wavefunction but without causing significant carrier trapping. Our approach allows for the direct examination of the implications of this unusual scattering mechanism for the carrier dynamics in MAPbI$_3$ by simulations of very large system sizes. This is a necessary attribute of a real-space approach, where the diffusive limit (which can be detected by a pronounced plateau in the time derivative of the quantum mean-square displacement) must be reached before the amplitude of the electronic wave function reaches the simulation boundaries. Indeed, as described in the SI, we use simulation cell lengths on the order of 100 unit cells in our final calculations, far larger than the expected size of polarons in MAPbI$_3$.[16] Finite size studies illustrate that the mobilities extracted with such large cells yield converged results.

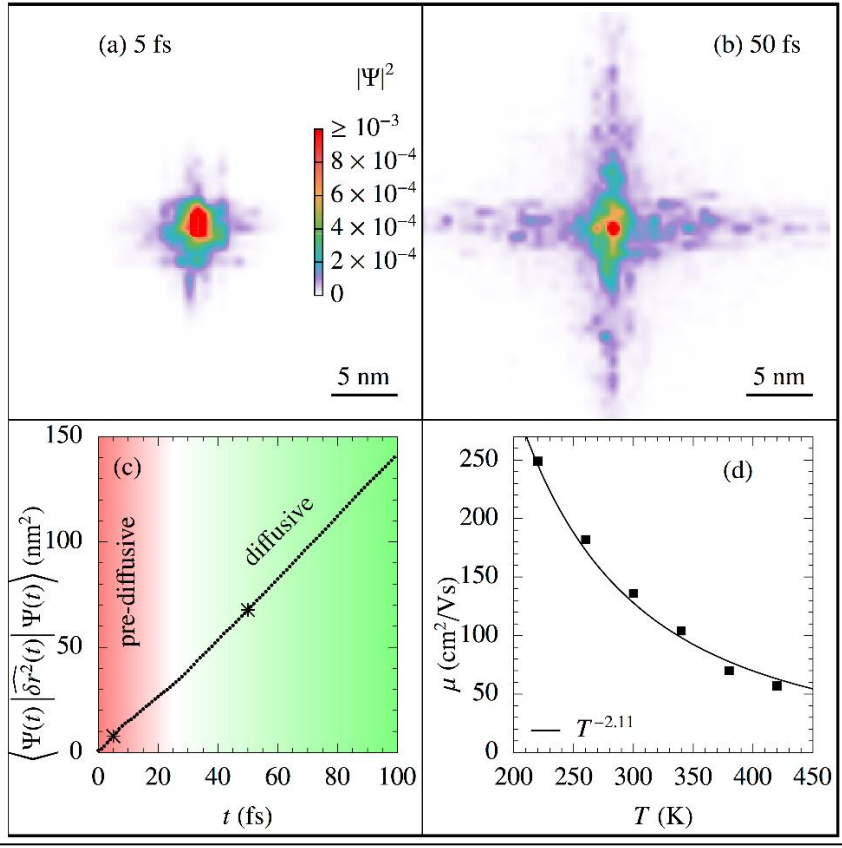

**Figure 2** (a) Two-dimensional projection of the spread of a conduction-band Wannier wave-packet (as described in the SI) at 5 fs (a) and at 50 fs (b). Note that while a localized wave-packet is used here for illustrative purposes, our mobility calculations employ the correct canonical initial condition. The wave-packet spread shows that the coherent transfer of free carriers is retarded by the disorder in the potential landscape and orbital fluctuations. The spreading is anisotropic due to fact that p-p Pb-I overlap dominates. Thus, it is expected that the mobility also encodes this anisotropy. (c) Calculated quantum mean-squared displacement from initial carrier position. Two time points corresponding to (a, pre-diffusive) and (b, diffusive) motion are illustrated. It shows that the orbital-overlap fluctuations result in nearly-free carriers that diffuse in MAPbI$_3$. The crossover time implies mean electron-phonon scattering times of 10fs or less at 300K. (d) Thermally averaged mobility (see SI) as function of temperature and a fit to the $T$-dependent data.

In Fig. 2d, we show the simulated $T$-dependence of electron mobilities. Remarkably, semi-quantitative agreement is found when the theory is compared to the results of recent experiments with respect to *both* the magnitude of the room-$T$ mobility (with a best estimate of $\mu_e \approx 135$ cm$^2$/Vs) as well as its $T$-dependence (approximately $\sim T^{-2}$).[23,34–36] The hole mobility is expected to be approximately 1.5 times smaller at 300 K (see SI). Experimental data on carrier mobilities in MAPbI$_3$ do scatter in the literature but are generally found to be of the order of, or slightly less than, 100 cm$^2$/Vs at room-$T$,[37] which is very close to our approximate result. Further improvements to our theory, such as using more advanced DFT functionals in the TB parameterization or including dynamic effects beyond displacements in the nuclear fluctuations may be required for being able to making even more accurate predictions of the carrier dynamics. It is also noted that the exponent in the *T*-dependence of the mobility has been reproduced in experiments to lie between ~-1.5 to –2.8[36,37], which is far above (300% to 500%) the exponent estimated from the large polaron model[16,17] but within 30% to 40% of our estimate.

Having found that our theoretical model allows for a largely accurate description of the carrier dynamics in MAPbI3 around room-*T*, our mobility data enable us to quantify the importance of the different scattering mechanisms included in our model. We find that the temporal and spatial fluctuations of orbital overlaps that arise from large atomic displacements and result in off-diagonal coupling fluctuations are the dominant factor in the *T*-dependence of the mobility (see SI). Our approach provides a fully microscopic theoretical description capable of capturing the non-trivial simultaneous consistency of the magnitude and *T*-dependence of the mobility. We have focused on the scattering mechanisms in HaPs pertinent around room-*T*, which is most important for their use in devices. But at lower *T* other mechanisms not included in our model, e.g., scattering by ionic impurities, could be important especially when the magnitude of nuclear displacements is reduced.

The theoretical framework outlined here may also be used to quantitatively describe and understand several puzzling features associated with the optical properties of HaPs. In particular, experiments have demonstrated that MAPbI3 exhibits a mild band gap *opening* with increasing temperature, as well as a sharp, weakly temperature-dependent absorption edge.[10,23–25,38] The former feature is difficult to rationalize with low-order electron-phonon coupling,[25] while the latter feature is surprising given the large scale of thermal disorder in MAPbI3 discussed above. Within our model, these puzzles may be addressed via calculation of the time-dependence of the instantaneous band gap at different temperatures driven by nuclear motion. Fig. 3a illustrates the temperature-dependent change in the band gap from a reference value at 300 K that is in very good agreement with experimental data. Previous theoretical work relied upon high-order electron-phonon coupling to describe the observed behaviour,[25] while such coupling, as well as anharmonic corrections, are naturally captured by our model. It should however be noted that we have neglected excitonic effects in our account of the band-gap opening. This is justified because the exciton binding energy only varies by ~10 meV from very low to room-*T*,[39] while the band gap

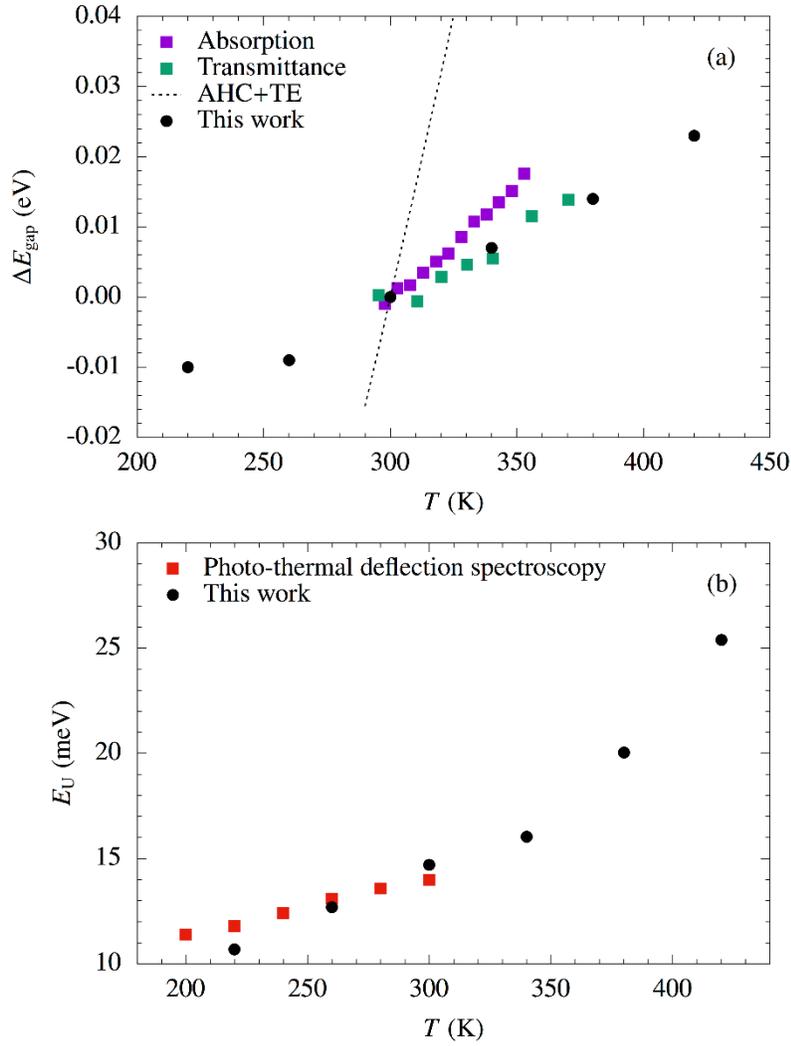

**Figure 3** (a) Experimental band gap as a function of temperature and theoretical data using linear-electron-phonon coupling (denoted "AHC+TE"), as adapted from Refs. 23-25, compared to theory of this paper; all data are aligned to their values at 300 K. (b) The Urbach energy quantifies the steepness of the optical absorption profile; comparison is made between the theory of this paper and the experimental results of Ref. 38. As charge and orbital fluctuations are correlated spatially only over short distances, the distribution of the $T$-dependent instantaneously recorded band gap values is narrow, resulting in small Urbach energies.

changes by ~20 meV over the much smaller $T$ range from 300 K to 350 K. As thermal nuclear motion also leads to a distribution of band gaps which vary with $T$,[40,41] it can affect the sharpness of the absorption edge as characterized by the Urbach energy, $E_U$.[42,43] Using the $T$-dependent width of the gap distribution (see SI for convergence tests), we find Urbach energies $E_U$= 10.7 meV (220 K) and 14.7 meV (300 K) as reported in Fig. 3b, which compare remarkably well with measured values of $E_U$=11.0 meV (200 K), 14.5 meV (300 K).[10,38] It should be noted that the off-diagonal orbital overlap fluctuations contribute approximately 50% to the magnitude of $E_U$ in our calculations.

The seemingly surprising confluence of a strongly limited mobility and a small $E_U$ may be understood by noting that charge scattering is completely determined by *local* lattice fluctuations, while the band gap is a *global* quantity that exhibits statistical self-averaging. Naively one expects that in an infinite size system band-gap fluctuations tend to zero, but this is only true when no spatial correlations of such fluctuations exist.[40] Instead, $E_U$ is known to depend on the correlation length of the disorder potential, where long-range correlations result in wider band-gap distributions and larger $E_U$.[41] We quantify the correlation length of the disorder potential to be on the order of less than one unit cell of $MAPbI_3$ (see SI), which shows that it is short-ranged. This explains the mild effect of the lattice fluctuations on the sharpness of optical absorption.

In conclusion, we have presented a unified, fully microscopic model capable of accurately describing several of the most important and perplexing properties of HaPs, including the unique features that limit charge transport and the sharpness of the absorption edge in these systems. Our approach combines a computationally efficient DFT-TB scheme with large-scale MD to allow for quantum-dynamical simulations of the prototype material $MAPbI_3$ that naturally incorporate the effects of the anharmonicity of nuclear motion and non-linear electron-phonon coupling. A major finding of our work is that large-amplitude anharmonic displacements in the lead-halide lattice involve charge and orbital fluctuations that lead to off-diagonal coupling modulations and play a leading role in determining these features, in stark contrast to typical inorganic semiconductors. In addition to providing a microscopic rationalization of the highly unusual properties of systems such as $MAPbI_3$, the theoretical framework presented here promises to enable the rational design of as yet undiscovered related materials with tailored and optimized optoelectronic properties.

**Acknowledgements**

We would like to thank David Cahen (Weizmann Institute of Science), Omer Yaffe (Weizmann Institute of Science), and Jorge Andrés Guerra (Pontificia Universidad Católica del Perú) for discussions. L.Z.T. acknowledges support from the Office of Naval Research under grant number N00014-17-1-2574. Funding provided by the National Science Foundation via grants No. CHE-1839464 (D.R.R), No. DMR-1719353 (A.M.R) and an NSF Gradudate Fellowship (M.Z.M.), and the Alexander von Humboldt Foundation in the framework of the Sofja Kovalevskaja Award endowed by the German Federal Ministry of Education and Research (D.A.E.), is acknowledged. The authors further acknowledge computational support from the High-Performance Computing Modernization Office of the Department of Defense, and computing time granted by the John von Neumann Institute for Computing (NIC) and provided on the supercomputer JURECA[44] at Jülich Supercomputing Centre (JSC) is appreciated as well.

## Author contributions

M.Z.R. performed the quantum dynamics simulations and analyzed the data. M.Z.R. and D.R.R. developed the theoretical models for the quantum dynamics simulations. L.Z.R. performed DFT calculations, parametrized the TB model, and analyzed the data. L.Z.R. and A.M.R. developed the TB model. D.A.E. performed the MD calculations. D.A.E., A.M.R., and D.R.R. conceived the study and wrote the manuscript. All authors discussed the results and contributed to the writing of the manuscript.